\documentclass{WileyMSP-template}
\usepackage{dcolumn}
\usepackage{graphicx}
\usepackage{epstopdf}
\usepackage{mathrsfs}
\usepackage{subfigure}
\usepackage{booktabs}
\usepackage{amsmath}
\usepackage{physics}
\usepackage{dsfont}
\usepackage{amstext}
\usepackage{amssymb}
\usepackage{amsbsy}
\usepackage{bbm}
\usepackage{amsthm}
\usepackage{xcolor}
\usepackage{url}

\usepackage{cite}

\usepackage[
    colorlinks=true,
    urlcolor=blue,
    linkcolor=blue,
    citecolor=blue,
    breaklinks=true,
    hypertexnames=false,
    pageanchor=true,
    anchorcolor=black
]{hyperref}

\begin{document}
\pagestyle{fancy}
\rhead{\includegraphics[width=2.5cm]{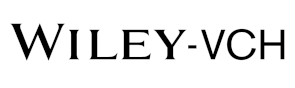}}
\title{Search for Axionlike Dark Matter by Quantum Technology}

\maketitle

\author{Jing Dong}
\author{Rong-Hang Chen}
\author{Qing Ai*}

\dedication{}

\begin{affiliations}
\author{Jing Dong}\\
School of Physics and Astronomy, Beijing Normal University, and Key Laboratory of Multiscale Spin Physics (Beijing Normal University), Ministry of Education, Beijing 100875, China\\
Institute of Physics, Chinese Academy of Sciences, Beijing 100190, China\\
University of Chinese Academy of Sciences, Beijing 100049, China\\
\author{Rong-Hang Chen}\\
Beijing Computational Science Research Center, Beijing 100193, China\\
School of Physics and Astronomy, Beijing Normal University, and Key Laboratory of Multiscale Spin Physics (Beijing Normal University), Ministry of Education, Beijing 100875, China\\
\author{Qing Ai*}\\
Email: aiqing@bnu.edu.cn\\
School of Physics and Astronomy, Beijing Normal University, and Key Laboratory of Multiscale Spin Physics (Beijing Normal University), Ministry of Education, Beijing 100875, China\\
\end{affiliations}


\begin{abstract}
%
The axion is a highly motivated elementary particle that could address two fundamental questions in physics— the strong charge--parity (CP) problem and the dark matter mystery. Quantum sensors offer complementary ways to search for its weak interactions through electromagnetic conversion, spin precession, and oscillating electric dipole moments. This review connects these mechanisms with quantum parameter estimation, coherent control, and realistic measurement noise. Squeezed microwave receivers, superconducting photon counters, spin ensembles, and distributed sensors are compared in terms of coupling sensitivity, coherence, bandwidth, and experimental maturity. Direct axion searches are distinguished from scalar-field spectroscopy, dark-photon measurements, and laboratory force constraints. The discussion emphasizes that quantum enhancement must survive loss, decoherence, thermal backgrounds, and experimental overhead to improve a search. Established demonstrations and proposed protocols are assessed separately, identifying low-loss transduction, calibrated quantum readout, and robust stochastic-signal analysis as priorities for future experiments.

\end{abstract}


\section{Introduction}

The microscopic nature of dark matter remains an open question despite the evidence for its gravitational effects on astrophysical and cosmological scales~\cite{Bertone2018RevModPhys,Clowe2006ApJL,Planck2020Parameters}. The axion offers a particularly compelling connection between this problem and the structure of the strong interaction~\cite{Kim2010RMP,DiLuzio2020PhysRep}. Introduced through the Peccei--Quinn mechanism, the quantum chromodynamics (QCD) axion dynamically suppresses strong charge-parity ($CP$) violation~\cite{Peccei1977PRL,Peccei1977PRD,Weinberg1978PRL,Wilczek1978PRL} and can also constitute dark matter through nonthermal production in the early Universe~\cite{Preskill1983PLB,Abbott1983PLB1,Dine1983PLB,Marsh2016PhysRep}. Axion-like particles (ALPs) extend the experimental target to light pseudoscalars whose masses and couplings need not obey the conventional QCD axion relation~\cite{Peter2006JHEP,Arias2012JCAP,Irastorza2018PPNP}. Searching for these particles therefore tests both a specific solution to the strong $CP$ problem, in the QCD axion case, and a broader class of weakly coupled dark-sector theories~\cite{Graham2015Annu,Sikivie2021RevModPhys}.

In the wave-like regime, axion dark matter can produce weak oscillatory perturbations that are approximately coherent over finite time and length scales, with stochastic amplitude and phase variations on longer scales~\cite{Graham2013PRD,Marsh2016PhysRep,Centers2021NC,yan2025prl}. Its interactions with photons and spins provide complementary ways to convert these perturbations into measurable signals~\cite{Graham2015Annu,Sikivie2021RevModPhys}. Microwave cavity haloscopes search for electromagnetic excitations generated by axion--photon conversion in an applied magnetic field~\cite{Bradley2003RMP,Du2018PRL,Brubaker2017PRL}, whereas nuclear magnetic resonance and comagnetometer experiments probe axion-induced spin dynamics~\cite{Budker2014PRX,Abel2017PRX,wu2019prl,Aybas2021prl,Bloch2023NC,Yan2022PRL}. The small expected signals and the unknown axion mass make sensitivity and frequency coverage inseparable experimental challenges: a detector must distinguish a weak excitation from noise while exploring a potentially broad range of signal frequencies~\cite{Chaudhuri2019Arxiv}. The expanding range of proposed and demonstrated detector platforms, including lumped-element, dielectric, and plasma haloscopes, addresses complementary axion mass ranges~\cite{Irastorza2018PPNP,Ouellet2019PRL,Caldwell2017PRL,Lawson2019PRL,Brouwer2022PRD}.

Quantum sensing addresses this challenge by controlling how a signal is encoded in a probe and how the resulting state is measured~\cite{Degen2017RMP,Sushkov2023PRXQuantum}. Coherent spin evolution, quantum-limited microwave amplification, squeezed fields, and superconducting circuits provide distinct resources for detecting weak interactions~\cite{Clerk2010RMP,Degen2017RMP,chou2023arXiv}. However, It is important to distinguish a sensor that uses quantum degrees of freedom from a protocol that demonstrably outperforms an appropriate uncorrelated-probe or conventional-readout benchmark under comparable resource constraints~\cite{Giovannetti2006PRL,Giovannetti2011NatPhoton,Pezze2018RMP}. Quantum enhancement is not a universal reduction of every noise source; its value depends on the measured observable, the signal statistics, and the physical processes that limit the measurement~\cite{Clerk2010RMP,Centers2021NC,Aybas2021QST,Kuo2025PRD}. This distinction is central to assessing quantum technologies for axion searches.

The approaches considered in this review illustrate several ways in which quantum control can improve the search. Squeezed-vacuum receivers suppress fluctuations in the measured microwave quadrature and can broaden the useful detection bandwidth, thereby increasing the scan rate for a signal of unknown frequency~\cite{Zheng2016Arxiv,Malnou2020PRX}. Engineered interactions between a sensing cavity and an auxiliary readout mode offer a complementary route to bandwidth enhancement, developed in theoretical proposals and demonstrated in weak-signal searches~\cite{Wurtz2021PRXquantum,Jiang2023PRXQuantum}. Microwave photon counting changes the readout observable from field quadratures to discrete excitations, with its advantage determined by thermal occupation, detection efficiency, and dark counts~\cite{Lamoreaux2013PRD,Kuo2025PRD}; this approach has now been implemented in an axion haloscope search~\cite{Braggio2025PRX}. In spin-based searches, resonant amplification and Floquet control improve the transduction and readout of weak axion-induced perturbations~\cite{Jiang2021NP,Jiang2022PRL,Bloch2022SA,Wei2025RPP}. These methods should be compared through their effects on detectable coupling strength, useful bandwidth, and acquisition time, rather than through signal amplification alone~\cite{Aybas2021QST,Kuo2025PRD}.

A common parameter-estimation framework helps connect these otherwise different platforms. Classical and quantum Fisher information characterize the information available in measurement outcomes and probe states, respectively, and provide a basis for discussing standard quantum limits and possible gains from squeezing or entanglement~\cite{Braunstein1994PRL,Paris2009IJQI,Giovannetti2006PRL,Pezze2018RMP}. The realizable gain nevertheless depends on decoherence, photon loss, thermal noise, and imperfect state preparation and readout~\cite{Huelga1997PRL,Escher2011NatPhys,Demkowicz2012NatCommun,Degen2017RMP}. Consequently, an improvement in an idealized precision bound does not by itself establish an improvement in an experimental dark-matter search: the signal statistics, accessible measurements, and search protocol must also be specified~\cite{Centers2021NC,Kuo2025PRD}. The sensor, coupling network, measurement protocol, and noise environment must be evaluated together~\cite{Clerk2010RMP,Aybas2021QST,Sushkov2023PRXQuantum}. Optimized quantum sensor networks extend this perspective to multiple probes~\cite{Zhang2021QST,Brady2022PRXQuantum,Santoso2025PRD}, but their proposed entanglement-enabled advantages must be distinguished from the performance already demonstrated by geographically distributed magnetometer networks~\cite{Pospelov2013PRL,Afach2021NATURE}.

This review examines axion dark-matter detection from the combined perspectives of particle interactions, quantum parameter estimation, and experimental readout. We first summarize the strong $CP$ problem, the Peccei--Quinn mechanism, and the axion couplings relevant to detection. We then introduce the quantum estimation framework and the roles of measurement limits, decoherence, and loss. Against this background, we discuss quantum-limited amplifiers, squeezed-state receivers, spin-ensemble techniques, and single-photon detection, emphasizing the conditions under which each approach improves a search. The final discussion considers lower-noise receivers, squeezing and entanglement, distributed sensor networks, and hybrid quantum systems. Axions and ALPs remain the principal targets; related dark-sector sensing schemes, including superconducting-qubit and nonclassical-state searches for dark photons, are considered where they clarify transferable measurement strategies~\cite{Dixit2021PRL,Agrawal2024PRL,Zheng2026PRL}.

\section{Theoretical Background of Axion Dark Matter}\label{sec:theory}
The strong interaction permits a vacuum angle whose effects are tightly constrained by experiment. The smallness of this angle motivates the QCD axion. The axion mass follows from the curvature of the vacuum energy, while a detector responds to a particular coupling and to the local field seen by its probe. We develop these ingredients in the order in which they enter a predicted signal.

\subsection{The strong CP problem and the Peccei–Quinn mechanism}
In QCD a dimension-four term consistent with gauge invariance can violate $CP$. Before the quark masses are made real by chiral field redefinitions, the gluonic term in the sign convention of Equation~\ref{eq:coupling_operators} is
\begin{equation}
\mathcal L_\theta=-\theta\frac{g_s^2}{32\pi^2}
G^A_{\mu\nu}\widetilde G^{A\mu\nu}.
\label{eq:theta_term}
\end{equation}
Here $g_s^2=4\pi\alpha_s$, $G^A_{\mu\nu}$ is the gluon field strength and $\widetilde G^{A\mu\nu}=\epsilon^{\mu\nu\rho\sigma}G^A_{\rho\sigma}/2$ its dual. The density $G\widetilde G$ is parity odd and charge-conjugation even. Its space-time integral distinguishes sectors of different topological charge, whose amplitudes acquire relative phases set by $\theta$ \cite{Callan1976PLB}. The chiral rotations that remove the quark-mass phases also shift this coefficient. In the mass-phase convention used here the physical combination is $\bar\theta=\theta+\arg\det(M_uM_d)$ modulo $2\pi$, with $M_u$ and $M_d$ the up- and down-type quark mass matrices \cite{Peccei1977PRD,Kim2010RMP}.

A nonzero $\bar\theta$ induces a permanent neutron electric dipole moment (EDM). A QCD sum-rule estimate, with the normalization corrected by its authors, gives \cite{Pospelov1999PRL}
\begin{equation}
|d_n^{(\bar\theta)}|\simeq 2.4\times10^{-16}\,|\bar\theta|\,e\,\mathrm{cm}.
\label{eq:neutron_edm}
\end{equation}
The coefficient has a hadronic uncertainty of roughly $40$--$50\%$. Building on the earlier search by Baker et al. \cite{Baker2006PRL}, Abel et al. obtained the experimental upper bound \cite{Abel2020PRL}
\begin{equation}
|d_n|<1.8\times10^{-26}\,e\,\mathrm{cm}\qquad (90\%\ \mathrm{C.L.}).
\end{equation}
In the absence of a cancellation with other EDM contributions, these results imply the order-of-magnitude constraint
\begin{equation}
|\bar\theta|\lesssim10^{-10}.
\label{eq:theta_bound}
\end{equation}
The Standard Model admits a vacuum angle this small but supplies no symmetry that explains the required combination of independent parameters. This naturalness question is the strong $CP$ problem \cite{Kim1987PR,Cheng1988PR,Kim2010RMP}.

Peccei and Quinn introduced a global chiral $U(1)_{\rm PQ}$ symmetry with a color anomaly \cite{Peccei1977PRL,Peccei1977PRD}. Its spontaneous breaking leaves an angular degree of freedom, the pseudoscalar axion $a_{\rm PQ}$ \cite{Weinberg1978PRL,Wilczek1978PRL}. With the anomaly coefficient absorbed into the decay constant $f_a$, QCD depends on the angle $\theta_{\rm eff}=\bar\theta+a_{\rm PQ}/f_a$. Nonperturbative QCD generates a periodic potential for this angle. For positive quark masses the vacuum energy is minimized at $\theta_{\rm eff}=0$ modulo $2\pi$ \cite{Vafa1984PRL,Grilli2016JHEP}. Its gradient exerts a restoring force; if additional sources of explicit PQ breaking are negligible, the equilibrium field cancels the static angle. The angular mode gains mass from the anomaly and is a pseudo-Nambu--Goldstone boson.

Near this minimum the zero-temperature topological susceptibility $\chi_{\rm QCD}$ determines the potential for the canonical field displacement $a=f_a\theta_{\rm eff}$,
\[
\chi_{\rm QCD}=\left.\frac{\partial^2 E_{\rm QCD}}{\partial\theta_{\rm eff}^2}\right|_0,\qquad
V(a)-V(0)=\tfrac12 m_a^2 a^2+O(a^4),\qquad m_a^2 f_a^2=\chi_{\rm QCD}.
\]
Here $E_{\rm QCD}$ is the vacuum energy density for a fixed angle. The curvature fixes the mass even though the global shape of the QCD potential differs from a pure cosine \cite{Grilli2016JHEP,Borsanyi2016Nature}.

\subsection{QCD axions and axion-like particles}
The QCD axion has a correlated mass and decay constant, whereas a generic axion-like particle (ALP) need not
obey this relation. Likewise, axion–photon and axion–fermion couplings are model dependent even at fixed $f_a$. An exclusion of one coupling over a narrow mass interval therefore does not exclude all ALP models or all QCD-axion realizations \cite{Kim2010RMP,Graham2015Annu}. In the following, $a$ denotes the displacement from the vacuum minimum: $a_{\mathrm{old}}+f_a\bar\theta\to a$. The static angle is canceled, and the oscillating effective angle is $a/f_a$. A displaced initial field can produce a cold population through vacuum misalignment. Whether PQ breaking precedes or follows inflation controls the spatial initial conditions and the relevance of isocurvature, strings, and domain walls. The abundance then depends on the thermal history and, where relevant, defect evolution. A detector should consequently state the assumed axion density rather than identify every mass with a unique cosmological abundance. High occupation number permits a classical stochastic-field description of virialized ultralight dark matter,
\begin{equation}
a(t,\mathbf x)\simeq a_0\cos(\omega_a t-\mathbf k_a\cdot\mathbf x+\phi).
\label{eq:halo_field}
\end{equation}
where the field amplitude $a_0$ can be estimated by the
Galactic dark-matter energy density 
\begin{equation}
\rho_a=\frac{m_a^2a_0^2}{2}.
\end{equation}

Equation~\ref{eq:halo_field} fixes the amplitude for a single mode. For a virialized halo, modes of different velocities interfere, producing a stochastic field with finite spatial and temporal coherence \cite{Foster2018PRD,Centers2021NC}. Its characteristic energy density $\rho_a$ can be a fraction of the total local dark-matter density. Section~\ref{sec:frequency} gives the coherence scales that enter the choice of measurement protocol.

One input to this extrapolation is the amount of string available to radiate. Figure~\ref{fig:pub_string_density} shows how Buschmann et al.\ inferred it in an adaptive-mesh simulation. The radial PQ mode has mass $m_r$, so $m_r/H$ measures the ratio of the Hubble radius to the string-core scale \cite{Buschmann2022NC}.

\begin{figure}
\centering
\medskip
\includegraphics[width=15cm]{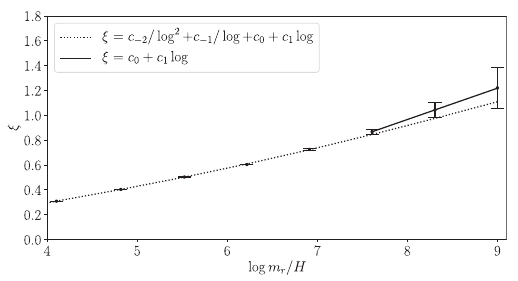}
\caption{String length in a postinflationary PQ simulation. The dimensionless density $\xi=\ell t^2/\mathcal V$ is obtained from the total string length $\ell$ in physical volume $\mathcal V$ at cosmic time $t$. Points show the simulation estimates, with fitted $68\%$ intervals. The dotted curve includes inverse powers of $\ln(m_r/H)$ over the full range. The solid line fits only the late-time points. Both give a leading slope near $0.25$. Reproduced from \href{https://doi.org/10.1038/s41467-022-28669-y}{Fig.~2} of Buschmann et al.\ \cite{Buschmann2022NC}, under \href{https://creativecommons.org/licenses/by/4.0/}{CC BY 4.0}.}
\end{figure}

\subsection{Main axion couplings relevant to experiments}
\begin{figure}[!htbp]
\centering
\includegraphics[width=18cm]{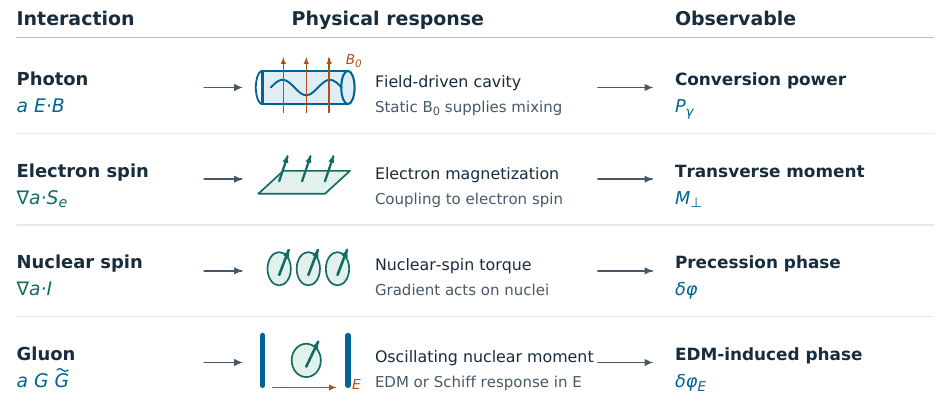}
\caption{Interaction channels for a coherent axion field. The photon coupling converts the field in a static magnet into a microwave signal. A spatial gradient acts separately on electron and nuclear spins through their derivative couplings; the resulting magnetization can be measured directly or converted into microwave power. The gluon coupling induces oscillating hadronic electric dipole moments or nuclear Schiff moments and can produce nuclear precession in an electric-field environment. Arrows identify transduction steps, independent of the relative reach of the detectors. Original schematic based on Refs.~\cite{Sikivie1983PRL,Graham2013PRD,Budker2014PRX,Aybas2021prl,Crescini2020PRL}.}
\label{fig:channels}
\end{figure}

A useful low-energy convention is
\begin{equation}
\mathcal L_{\mathrm{int}}=-\frac{g_{a\gamma}}4aF_{\mu\nu}\widetilde F^{\mu\nu}
+\sum_f\frac{C_f}{2f_a}(\partial_\mu a)\bar\psi_f\gamma^\mu\gamma^5\psi_f
-\frac{\alpha_s}{8\pi}\frac{a}{f_a}G^A_{\mu\nu}\widetilde G^{A\mu\nu}.
\label{eq:coupling_operators}
\end{equation}
The photon coupling $g_{a\gamma}$ has dimensions of inverse energy. The derivative interaction is written with dimensionless coefficients $C_f$; a commonly used pseudoscalar convention is $g_{af}=C_fm_f/f_a$ for fermion mass $m_f$. At nuclear energies the quark and gluon operators are matched onto nucleon couplings and nuclear matrix elements \cite{Graham2013PRD,Grilli2016JHEP}. The last term in Equation~\ref{eq:coupling_operators} uses the QCD axion normalization of $f_a$. An independent gluonic coefficient is possible for a general ALP. Figure~\ref{fig:channels} maps these interactions to the quantity actually collected by an apparatus.

In a static magnetic field $\mathbf B_0$, the photon operator sources an oscillating electromagnetic response proportional to $g_{a\gamma}\dot a\mathbf B_0$. A resonant haloscope converts this drive into microwave power. Suppressing apparatus-dependent unit factors, the resonant power scales as
\begin{equation}
P_a\propto g_{a\gamma}^2\frac{\rho_a}{m_a}B_0^2VC\min(Q_L,Q_a)\frac{\beta}{1+\beta},
\label{eq:haloscope_scaling}
\end{equation}
where $V$ is conversion volume, $C$ a normalized mode-overlap factor, $Q_L$ the loaded quality factor, $Q_a$ the halo quality factor, and $\beta$ the readout coupling. The minimum prescription is an asymptotic overlap approximation; quantitative calculations must integrate the actual halo spectrum against the resonator response. This scaling is not a replacement for a calibrated, unit-consistent apparatus response \cite{Graham2015Annu}.

The derivative interaction gives a nonrelativistic spin Hamiltonian proportional to $-(C_f/f_a)\mathbf {S}\cdot\mathbf {\nabla} a$. The axion gradient therefore behaves as a spin-dependent pseudomagnetic field. Its directional modulation and species dependence distinguish it from an ordinary magnetic background. The gluonic operator instead produces oscillating hadronic EDMs, detectable through electric-field-induced spin torque, nuclear screening and material response must be included. These routes are complementary measurements of distinct operators.

Scalar fields can modulate masses or the fine-structure constant and thereby shift clock-frequency ratios. Dark photons couple through vector kinetic mixing and induce electromagnetic fields with different shielding and polarization responses. Finally, a polarized laboratory source can probe an exchanged axion-mediated force without any halo population. Such experiments are valuable comparators, but their parameter limits cannot be relabeled as direct axion-dark-matter constraints. 
\begin{figure}
\centering
\includegraphics[width=15cm]{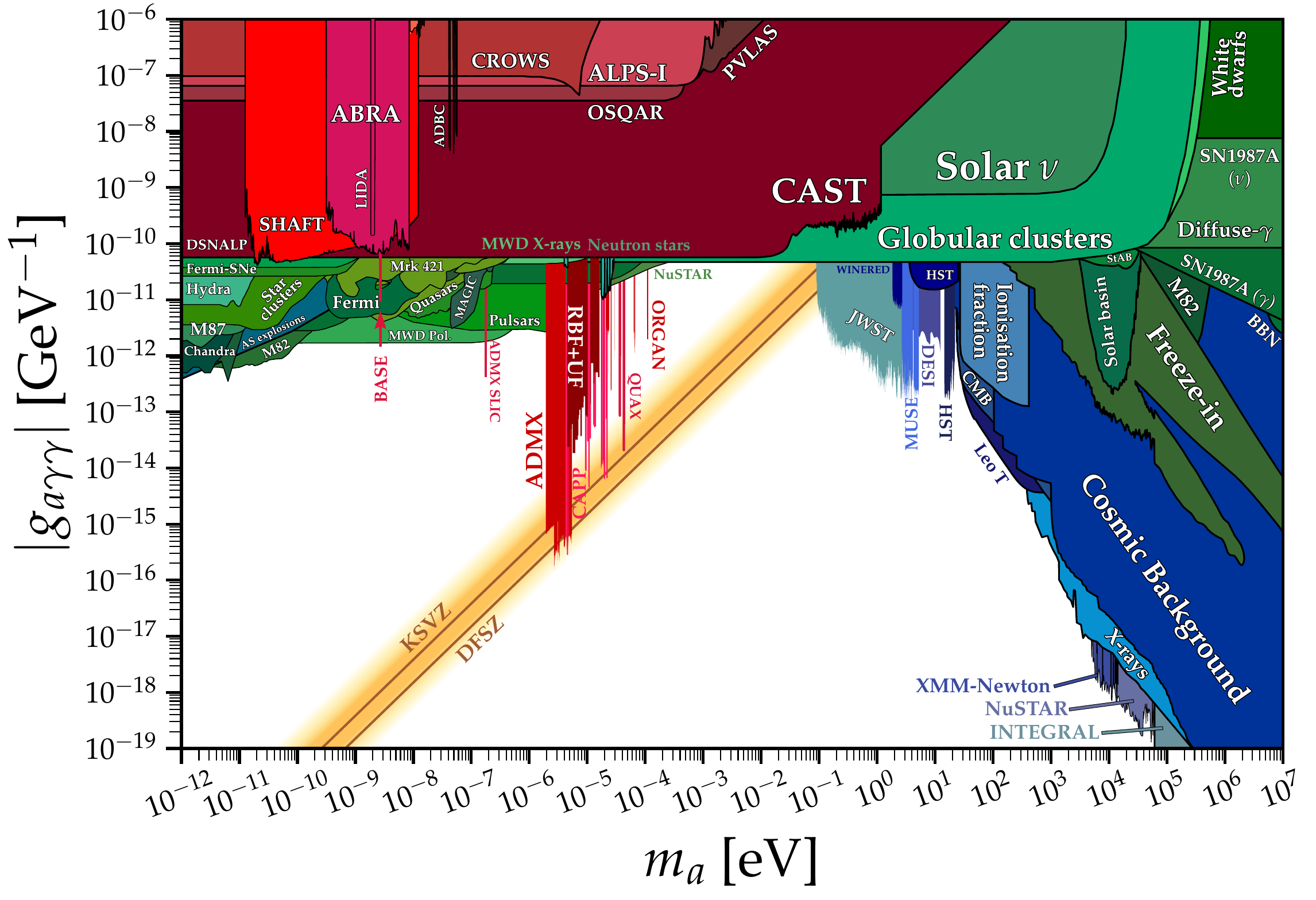}
\caption{Absolute axion--photon coupling as a function of axion mass. The AxionLimits compilation includes laboratory and astrophysical constraints and KSVZ and DFSZ QCD axion benchmarks Ref. \cite{Hare}}
\end{figure}

\subsection{Axial mass range and experimental frequency band}\label{sec:frequency}
Restoring constants, the leading oscillation frequency is $\nu_a=m_ac^2/h$, whereas $\omega_a=m_ac^2/\hbar$ is the angular frequency. Numerically, $1\,\mu\mathrm{eV}/c^2$ corresponds to approximately $241.8\,\mathrm{MHz}$. We reserve $f_a$ for the decay constant and use $\nu_a$ for frequency. For characteristic halo speed $v\sim10^{-3}$ in natural units, the fractional linewidth is of order $v^2$, the coherence time is of order $(m_av^2)^{-1}$, and the coherence length is of order $(m_av)^{-1}$; precise factors of $2\pi$ depend on the adopted definitions \cite{Graham2013PRD,Foster2018PRD}. Streams and transient defects require different correlation models.

Frequency, instantaneous bandwidth, and scanned coverage are separate properties. As two examples, a superconducting LC circuit in the BASE experiment constrained $g_{a\gamma}$ near $m_a=2.79\,\mathrm{neV}/c^2$, corresponding to about $674 \mathrm{kHz}$ \cite{Devlin2021PRL}, whereas QUAX scanned a $38\,\mathrm{MHz}$ interval around $10.2\mathrm{GHz}$, near $42 \mu\mathrm{eV}/c^2$ \cite{Infirri2025PRL}. Neither measurement covers an entire axion production scenario. At lower frequencies, spin precession and clock comparisons offer longer interrogation times, but the usable response depends on relaxation, control, and readout. These physical constraints motivate the platform-based organization of the next section.

Two proposed electromagnetic searches illustrate how the resonance can be matched to different parts of this frequency range. DMRadio-GUT would use a tunable lumped-element receiver at frequencies where the apparatus is small compared with the electromagnetic wavelength \cite{Brouwer2022PRD}. At higher frequencies, Lawson et al. proposed tuning the plasma frequency of a wire medium to the axion mass, allowing resonant conversion without setting the transverse size of the magnetized volume by the vacuum photon wavelength \cite{Lawson2019PRL}.

\section{Quantum Framework for Axion Detection}\label{sec:platforms}
The interaction channel determines what the detector must convert into a measurable observable. We therefore organize platforms by the degrees of freedom that carry the signal and the method used to read them out. The comparisons below concern detection mechanisms and demonstrated operating regimes; quantitative criteria for quantum enhancement are developed separately in Section~\ref{sec:enhancement}.

\subsection{spin ensembles}
Spin ensembles detect a weak drive through accumulated phase, transverse magnetization, or resonant transitions. Nuclear spins provide long-lived probes of axion--nucleon interactions and oscillating EDMs, while electron spins couple to the axion--electron interaction and often allow rapid optical preparation and readout \cite{Budker2014PRX,wu2019prl,Aybas2021prl,Bloch2023NC}. In a comagnetometer, species with different magnetic and exotic couplings help discriminate the desired response from ordinary magnetic perturbations. The discrimination depends on the calibrated transfer functions and nuclear spin content, rather than on perfect magnetic immunity.  Two implementations of the Cosmic Axion Spin Precession Experiment (CASPEr) illustrate how nuclear magnetic resonance (NMR) selects the interaction. Garcon et al. used ultralow-field NMR to search for the oscillating effective magnetic field produced by an axion gradient \cite{Garcon2019SA}. Aybas et al. used a polarized ferroelectric sample, in which an oscillating nuclear Schiff moment can drive resonant spin precession in the internal electric field \cite{Aybas2021prl}. Both experiments read out nuclear magnetization. Translating that observable into a coupling requires the nuclear spin content for the wind channel and the nuclear and material responses for the electric channel.

For a nucleus small compared with the axion wavelength, the same gradient acts on all its nucleons. At leading order the spin-dependent perturbation within a fixed nuclear multiplet follows by projecting the nucleon-spin operators,
\[
H_{\nabla}^{\rm nuc}=-\frac{1}{f_a}\,\nabla a\cdot
 \left(C_p\mathbf S_p+C_n\mathbf S_n\right).
\]
Here $C_p$ and $C_n$ are the matched proton and neutron couplings, and $\mathbf S_p$ and $\mathbf S_n$ are the respective total spin operators restricted to the nuclear states used. Matrix elements of the spin component transverse to the bias field set the resonant transition strength \cite{Graham2013PRD,Budker2014PRX,Garcon2019SA}. For a comagnetometer, the inferred signal further depends on the magnetic response of both species and on their relative orientation to the axion gradient.

The transverse relaxation time $T_2$ sets the intrinsic width of a homogeneous spin resonance. The frequency interval over which a weak drive can be detected also depends on which noise sources follow the spin susceptibility. In a $^{21}$Ne search, Xu et al. measured both the resonant response and the noise and used frequency steps about thirty times the resonance width to constrain ALP--nucleon couplings between $4.5$ and $15.5\,\mathrm{Hz}$ \cite{Xu2024CP}. The halo coherence time enters a different part of this analysis. It sets the duration over which a phase realization of the incident field persists. For a gradient coupling, both the amplitude and the direction fluctuate, so a likelihood for short data records must include the distribution of the gradient projected onto the sensitive axis \cite{Lisanti2021PRD}.

\begin{figure}[!htbp]
\centering
\includegraphics[width=18cm]{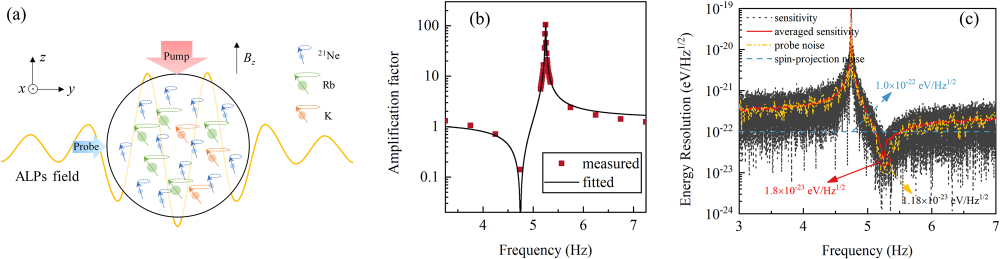}
\caption{Calibration of the K--Rb--$^{21}$Ne spin sensor. (a) Optical pumping and probing of the coupled electrons and nuclei. (b) Measured amplification of a 1 pT oscillatory transverse magnetic field and its fit. (c) Noise expressed as equivalent spin-energy amplitude per square root bandwidth at a bias setting with resonance frequency $5.25\,\mathrm{Hz}$. The red curve averages over $0.07\,\mathrm{Hz}$. Reproduced from \href{https://doi.org/10.1038/s42005-024-01713-7}{Fig.~4} of Xu et al. \cite{Xu2024CP} under \href{https://creativecommons.org/licenses/by/4.0/}{CC BY 4.0}.}
\label{fig:pub_xu_spin}
\end{figure}

The ChangE-NMR calibration illustrates why the width of the spin resonance alone cannot fix the scan spacing. Figure~\ref{fig:pub_xu_spin} compares the response to a known transverse magnetic drive with the energy resolution inferred from the optical record \cite{Xu2024CP}.

Coupling nuclear spins to alkali-metal spins makes weak nuclear responses optically accessible. In NASDUCK, polarized $^{129}$Xe nuclei were read out with a Floquet-driven $^{85}$Rb magnetometer, bridging the mismatch between nuclear and electronic resonances. The search covered $1$--$1000\,\mathrm{Hz}$ and established terrestrial ALP--neutron constraints \cite{Bloch2022SA}. A K--Rb--$^{21}$Ne implementation combined low-frequency self-compensation with hybrid spin resonance to search over $0.01$--$1000\,\mathrm{Hz}$ \cite{Wei2025RPP}. These are examples of control-assisted transduction and bandwidth engineering; their performance should not automatically be described as an entanglement advantage.

Solid-state spins provide a complementary architecture. Chigusa et al. proposed electron-spin searches with nitrogen-vacancy (NV) centers in diamond using Ramsey, echo, dynamical-decoupling, and resonant-transition protocols \cite{Chigusa2025JHEP}. A related proposal uses the embedded $^{14}$N nuclear spin and electron--nuclear control to probe combinations of proton and neutron couplings \cite{Chigusa2025PRD}. These are projected sensitivities, with different interaction matrix elements, coherence times, and readout efficiencies. Long nuclear coherence does not, by itself, imply a stronger coupling limit than an electron-spin measurement.

Macroscopic ensembles also require a quantitative account of intrinsic spin noise. Kuenstner et al. resolved quantum fluctuations in a solid-state nuclear ensemble containing approximately $5\times10^{21}$ spins using a superconducting resonant circuit and a superconducting quantum interference device (SQUID) \cite{Kuenstner2026NP}. This establishes access to spin-projection noise and motivates future nonclassical preparation; it is not itself an axion exclusion or a demonstration of spin squeezing. The associated resource question is taken up in Section~\ref{sec:states}.

Finally, the same source--sensor apparatus can test axion-mediated forces. Su et al. used separate polarized $^{129}$Xe ensembles and alkali-spin readout to constrain neutron--neutron pseudoscalar interactions, combining resonant amplification with matched filtering \cite{Su2024PRL}. Here the controlled source produces the perturbation through virtual exchange: the mediator mass sets the force range, and the result does not assume ambient axion dark matter. This distinction is essential for the bound-state proposal discussed in Section~\ref{sec:bound}, which builds on this type of apparatus.

\subsection{Clocks}
Clocks measure shifts in transition frequencies or frequency ratios. Their applicability to axions must be established through the operator that shifts the relevant atomic, molecular, or nuclear energy levels. For comparison, Kennedy et al. used a strontium optical clock, a silicon cavity, and a hydrogen maser to constrain ultralight scalar dark matter through variations of fundamental constants \cite{Kennedy2020PRL}. Molecular vibrational and rotational transitions provide related sensitivity to mass ratios and symmetry-violating interactions \cite{DeMille2024NP}. Such observations cannot be converted directly into a linear pseudoscalar axion coupling limit.

A QCD axion can affect clock frequencies through its gluonic coupling. Kim and Perez related the induced changes in hadronic parameters to a quadratic dependence of atomic transition energies on the axion field and used spectroscopy data to constrain this channel \cite{Kim2024PRD}. For a single harmonic field, this dependence produces a constant term and an oscillation at twice the Compton frequency. A velocity distribution also generates slower fluctuations of the field intensity, which can be sought in clock records \cite{Flambaum2023PRD}. Banerjee et al. identified a measurable nuclear contribution in a heavy ion. They used the different sensitivities of the electric-quadrupole and electric-octupole transitions in $^{171}\mathrm{Yb}^{+}$ to the nuclear charge radius to interpret a long-term comparison of the two frequencies \cite{Banerjee2025PRL}. The inferred axion constraint depends on both the nuclear response to the field and the electronic response to the radius.

\begin{figure}[!htbp]
\centering
\includegraphics[width=12cm]{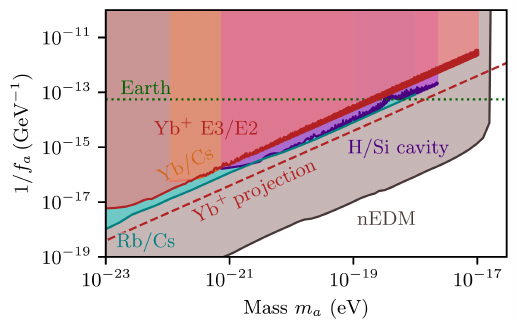}
\caption{Axion--gluon coupling scale inferred from the $^{171}\mathrm{Yb}^{+}$ optical frequency ratio. The dark-red boundary follows from the measured limit on oscillations of $\nu_{\rm E3}/\nu_{\rm E2}$; the dashed red line assumes a fractional modulation limit of $10^{-18}$. Other curves give spectroscopic and neutron-EDM constraints and a bound from the effect of terrestrial matter on the axion potential. Reproduced from \href{https://doi.org/10.1103/37vw-gc1r}{Fig.~2} of Banerjee et al.\ \cite{Banerjee2025PRL}, under \href{https://creativecommons.org/licenses/by/4.0/}{CC BY 4.0}.}
\label{fig:published_yb_limits}
\end{figure}

For a clock ratio $R=\nu_1/\nu_2$, let $K_R=\partial\ln R/\partial[(a/f_a)^2]|_{a=0}$ denote the differential response of these transitions to the axion-induced change of hadronic parameters. For a small single-mode field at the clock, Equation~\ref{eq:halo_field} gives the leading fractional shift \cite{Kim2024PRD,Banerjee2025PRL}
\[
\frac{\delta R(t)}{R}=
 K_R\frac{a_0^2}{2 f_a^2}
 \left[1+\cos(2\omega_a t+2\phi)\right].
\]
The phase $\phi$ includes the local spatial phase. A constant offset is degenerate with the unknown unperturbed ratio in a modulation search, which accesses the oscillatory part filtered by the interrogation sequence. Over many halo coherence times the single-mode form must be replaced by the statistics of $a^2$ \cite{Flambaum2023PRD}. The coefficient $K_R$ is specific to the transitions and to the assumed axion coupling. The clock-frequency ratio can constrain different interactions through different response coefficients. Filzinger et al.\ interrogated the two ytterbium-ion transitions in alternation and used their ratio to test for a variation of the fine-structure constant \cite{Filzinger2023PRL}. Banerjee et al.\ used the dependence of the same transitions on the nuclear charge radius to test the axion--gluon interaction. Figure~\ref{fig:published_yb_limits} shows the resulting inference from approximately 235 days of data acquired over 26 months \cite{Banerjee2025PRL}.

The clock record constrains a modulation amplitude. Its conversion to $1/f_a$ requires the differential nuclear and electronic response, an assumed local density (here $0.4\,\mathrm{GeV\,cm^{-3}}$), and the statistics of the field amplitude. At sufficiently low masses the observing span covers only part of a cycle. These dependencies determine how improvements in frequency resolution would move the red boundary \cite{Banerjee2025PRL}. The plot varies mass and gluonic coupling independently. Applying it to a particular QCD axion also requires the mass relation in Section~\ref{sec:theory}.

Interrogation protocols determine which frequency components are retained. Ramsey averaging suppresses sufficiently rapid oscillations, whereas echo and dynamical decoupling can reshape the response. Zaheer et al. developed narrowband and randomized broadband control protocols for scalar-dark-matter clock searches, including prospects for the $^{229}$Th nuclear transition \cite{Zaheer2025PRA}. Their analysis supplies a useful control strategy, but applying it to axions first requires the appropriate axion-induced clock Hamiltonian. Clock comparisons can then be extended across locations, making synchronization and spatial correlations the next relevant design questions.

\subsection{Distributed quantum sensor networks}
Distributed sensors can test correlations predicted by a spatially extended dark-matter field while rejecting local backgrounds. For a virialized field, the usable baseline and phase relation depend on the coherence length and sensor orientation; for a propagating defect, the relevant signature is a sequence of correlated transients \cite{Pospelov2013PRL,Afach2021NATURE,Foster2018PRD}. Classical combination of measurements from quantum sensors and distribution of entanglement between sensors are distinct architectures.

Gavilan-Martin et al. combined two K--Rb--$^3$He comagnetometers located in Mainz and Krak\'ow to search for ALP-gradient interactions, including the effect of changing sensor orientation with Earth's rotation \cite{Gavilan-Martin2025NC}. In a different signal model, Wang et al. used five noble-gas sensors in Hefei and Hangzhou, with an intercity baseline of about $320\,\mathrm{km}$, to search for axion topological-defect transients \cite{Wang2026Nature}. Their $739\,\mathrm{h}$ data set yielded no significant events. The reported reach, including $f_{\mathrm{int}}=4.1\times10^{10}\,\mathrm{GeV}$ near $84\,\mathrm{peV}/c^2$, is conditional on the adopted defect abundance and encounter model, and is not a generic virialized-halo bound.

A useful comparator is the 1692 km network of 15 atomic magnetometers used by Jiang et al. to constrain dark-photon kinetic mixing over 1--500 Hz \cite{Jiang2024NC}. It illustrates how cross-correlation can suppress uncorrelated backgrounds; its interaction and field geometry differ from those of an axion gradient. These distributed experiments demonstrate coordinated sensing, not entanglement shared between distant nodes.

Entangled electromagnetic sensor networks remain a separate development. Brady et al. proposed distributing squeezed microwave resources among cavities and coherently combining their outputs within a common dark-matter coherence region \cite{Brady2022PRXQuantum}. This architecture requires phase-compatible transduction and a sufficiently efficient distribution network. Its quantum resource and scan-rate comparisons are discussed in Section~\ref{sec:states}, while the next subsection considers how distinct subsystems can be combined within one detector.

\subsection{Hybrid quantum systems}
Hybrid detectors separate signal collection, coherent evolution, and readout among subsystems chosen for different strengths. An axion-induced electromagnetic flux may be collected by a resonator and measured with atomic spins or a superconducting circuit; an axion-induced spin response may instead be transferred to an optical or microwave observable. The key experimental quantity is the calibrated transfer from the target interaction to the final record, including conversion loss and added noise \cite{Clerk2010RMP,Degen2017RMP}.

In an inductive haloscope the static magnet supplies the axion--photon conversion field and an $LC$ circuit collects the oscillating magnetic flux, as proposed by Sikivie, Sullivan, and Tanner \cite{Sikivie2014LC}. Kim et al. used an alkali-vapor optical magnetometer as the readout. With a room-temperature pickup circuit and the unenergized magnet cooled to $4\,\mathrm K$, the prototype measured a background near $307\,\mathrm{kHz}$ of approximately $10\,\mathrm{fT}/\sqrt{\mathrm{Hz}}$. The projected cryogenic search assumes a sensor noise near $10\,\mathrm{aT}/\sqrt{\mathrm{Hz}}$ and targets QCD axions near $10^{-8}\,\mathrm{eV}/c^2$ \cite{Kim2023PRD}. In the ferrite--SQUID instrument of Kang et al. the electron spins in the magnetic material respond directly to the axion gradient; the experiment obtained sub-$100\,\mathrm{aT}/\sqrt{\mathrm{Hz}}$ sensitivity at tens of kilohertz and set axion--electron constraints over 5--60 kHz \cite{Kang2026PRL}. In the ferromagnetic haloscope of Crescini et al., illustrated schematically in Figure~\ref{fig:quantum_platforms}(d), an axion--electron interaction can excite a collective spin mode in yttrium iron garnet. The mode hybridizes with a microwave resonator. Tuning the bias field swept the resonance, and a Josephson parametric amplifier read out the resulting microwave power over $42.4$--$43.1\,\mu\mathrm{eV}/c^2$ \cite{Crescini2020PRL}. This identifies the spin matrix element, cavity coupling and readout efficiency that enter the interpretation of a power spectrum.

Interacting spin species also allow interference engineering. Jiang et al. observed amplification and deamplification arising from the bidirectional coupling of $^{87}$Rb and $^{129}$Xe, with the response described by Fano interference \cite{Jiang2025pnas}. The experiment demonstrates control of the magnetic transfer function, not an axion exclusion or unconditional noise suppression in every channel. At microwave frequencies, Chen et al. proposed coupled-mode haloscope arrays with matched beam-splitter and parametric interaction rates \cite{Chen2022PRR}. Their parity--time-symmetric response designs predict broader useful bandwidth, but require stability, control of internal loss, and compatibility with the conversion field. The associated quantum-readout mechanisms are assessed in Section~\ref{sec:receivers}.

Hybrid optical schemes can address laboratory production rather than halo detection. Dark spontaneous parametric down-conversion, proposed by Estrada et al., produces a visible signal photon together with an invisible boson under suitable phase matching \cite{Estrada2021PRXQuantum}. The weak-coupling dependence differs from production-and-regeneration experiments, but a favorable coupling power does not alone establish competitive sensitivity. We retain this proposal as a related use of quantum optics and distinguish it from the ambient-field platforms summarized in Table~\ref{tab:platforms}.

\begin{table}[!htbp]
\caption{Platform, interaction, and observable. Experimental searches and projected extensions are distinguished in the cited sections; using quantum degrees of freedom alone does not establish a quantum enhancement.}
\label{tab:platforms}
\centering\small
\begin{tabular}{p{0.16\textwidth}p{0.23\textwidth}p{0.23\textwidth}p{0.27\textwidth}}
\hline
Platform & Interaction or target & Recorded observable & Interpretation requirement\\
\hline
Electromagnetic haloscopes & Axion--photon conversion in a static field &  Cavity output power or pickup-loop flux & Conversion geometry, tuning response and calibrated noise\\
Spin ensembles & Axion--electron or axion--nucleon coupling; oscillating EDM & Spin phase, transverse polarization, optical rotation & Spin matrix elements, magnetic rejection, relaxation and control response\\
Clocks & Model-specific axion-induced energy shifts; scalar fields as comparators & Transition-frequency ratio or correlated fluctuations & Nuclear/atomic coefficients and linear versus quadratic response\\
Sensor networks & Coherent halo fields or specified transient defects & Cross-spectra, phase correlations, timed transients & Spatial coherence, orientation, correlated backgrounds; entanglement must be established separately\\
Hybrid systems & Photon conversion or spin coupling, depending on collector & Flux, optical signal, microwave quadrature or counts & Calibrated transduction, conversion efficiency, readout noise\\
Source--sensor force apparatus & Virtual axion exchange between polarized sources & Source-correlated spin response & Force geometry and mediator range; no halo-density inference\\
\hline
\end{tabular}
\end{table}

The platforms above establish where an interaction enters the apparatus and how it reaches the recorded signal. Improving that signal requires a second analysis: which noise or information loss limits the measurement, what resource can reduce it, and whether the gain survives the complete observing protocol. These questions define the scope of Section~\ref{sec:enhancement}.

\section{Quantum enhancement and Quantum measurement limits}\label{sec:enhancement}
For a specified coupling and halo model, the probe dynamics determine how strongly the field changes the probe state. The measurement then selects which changes appear in the data. A comparison between protocols therefore requires the same physical signal and a statement of the constrained resources, including conversion volume or spin number, observing time, and usable bandwidth. We develop that comparison first for phase accumulation and then for state protection, critical response, nonclassical probes, and microwave receivers.

\subsection{Axion detection as parameter estimation}
\label{sec:estimation}
At a trial mass, let $g$ denote the coupling of interest in a specified interaction convention. Preparation and evolution produce a state $\rho_g$, and an accessible measurement generates a distribution $p(x|g)$. The halo density, stochastic field correlations, detector transfer function, and instrumental backgrounds enter that distribution \cite{Foster2018PRD,Centers2021NC}. A spin phase, electromagnetic field amplitude, and photon excess therefore provide different encodings of the same underlying search problem.

In a Ramsey measurement, a first $\pi/2$ pulse prepares a spin superposition and a second pulse converts the accumulated relative phase into a population difference \cite{Ramsey1950PR}. For a weak longitudinal frequency shift $\delta\omega_a$, the signal phase in an interrogation starting at $t_0$ is
\[
\Phi(t_0)=\int_0^T y(t)\,\delta\omega_a(t_0+t)\,dt .
\]
Here $T$ is the free-evolution time and $y(t)=1$ between ideal Ramsey pulses. Inserted refocusing pulses can reverse the sign of $y(t)$, changing the band of oscillatory signals to which the sequence responds \cite{Degen2017RMP}. For fringe contrast $\mathcal C$ and analysis phase $\phi_{\mathrm r}$, one output has probability 
\begin{equation}
p_+=[1+\mathcal C\cos(\Phi-\phi_{\mathrm r})]/2.    
\end{equation}
At midfringe, projection noise in $N$ independent spins gives $\Delta\Phi=1/(\mathcal C\sqrt N)$ for one interrogation. The calibration from phase to coupling still requires the interaction strength, field statistics, and frequency response.

Schulthess et al. used separated oscillating fields in a cold-neutron beam to search for a time-dependent neutron electric dipole moment \cite{Schulthess2022PRL}. An applied electric field converts that moment into a precession-frequency shift. A continuous receiver instead records a filtered time series, from which a trial signal can be tested through its temporal correlations or power spectrum \cite{Clerk2010RMP,Foster2018PRD}. In either case the likelihood describes the accessible record. Ramsey windows, dead time, and fringe contrast enter one likelihood, while receiver bandwidth and added noise enter the other. For a stochastic halo field, combining records across coherence intervals requires a model for their correlations \cite{Centers2021NC}.

For $n$ independent, identically distributed repetitions, the single-repetition classical Fisher information (CFI) and the local Cram\'er--Rao bound are
\begin{equation}
F_C(g)=\sum_x p(x|g)[\partial_g\ln p(x|g)]^2,\qquad \mathrm{Var}(\hat g)\geq\frac{1}{nF_C(g)}.
\label{eq:fisher_bound}
\end{equation}
The sum becomes an integral for continuous outcomes. The variance bound assumes an unbiased estimator and the usual regularity conditions; correlated data require the joint likelihood rather than an assumed factor of $n$ \cite{Paris2009IJQI,Foster2018PRD}.

The quantum Fisher information (QFI) is $F_Q=\mathrm{Tr}(\rho_g L_g^2)$, where the symmetric logarithmic derivative satisfies $\partial_g\rho_g=(L_g\rho_g+\rho_gL_g)/2$. It bounds the CFI of any measurement, $F_C\leq F_Q$, yielding the quantum Cram\'er--Rao bound (QCRB), $\mathrm{Var}(\hat g)\geq1/(nF_Q)$, under the conditions of Equation~\ref{eq:fisher_bound}. For a parameter-independent pure input and unitary encoding $e^{-igtG}$, with a parameter-independent Hermitian generator $G$, $F_Q=4t^2\mathrm{Var}(G)$ \cite{Braunstein1994PRL,Paris2009IJQI}. The expression is not unchanged under dissipation or phase averaging. Even for one parameter, the locally optimal readout can depend on an unknown working point or require unavailable operations.

Unknown field phase, calibration, background, and resonance frequency can be nuisance parameters $\bf{\eta}$. In a regular classical model with an invertible Fisher matrix, the effective information about $g$ is $F_{gg}-F_{g\eta}F_{\eta\eta}^{-1}F_{\eta g}$. Stochastic halo variables must instead be treated consistently with their probability distribution \cite{Foster2018PRD,Cowan2011EPJC}. At the quantum level, optimal measurements for different parameters need not be compatible, so an inverse QFI matrix is not automatically an attainable covariance matrix \cite{Ragy2016PRA}. Phase-insensitive signals often first appear through $g^2$, making the sign unidentifiable and the zero-signal boundary nonregular. A coupling exclusion consequently requires a likelihood and coverage prescription, while a discovery claim across a frequency scan requires a trials correction and instrumental validation \cite{Cowan2011EPJC,Gross2010EPJC}.

Resource accounting makes a quantum benchmark meaningful. For $N$ independent identical probes with bounded single-probe generators, fixed interrogation time, and independent repetitions, additive information gives the standard quantum limit (SQL), $\Delta g\propto N^{-1/2}$. Suitable entangled inputs can yield an ideal local scaling $\Delta g\propto N^{-1}$ under the same unitary encoding \cite{Giovannetti2006PRL,Giovannetti2011NatPhoton,Pezze2018RMP}. Preparation, control, reset, and readout time must be included when comparing information per total observing time. Finite probe arrays approaching an ideal bound do not establish asymptotic scaling.

The amplifier SQL is a different benchmark: a high-gain phase-insensitive linear amplifier adds at least one-half quantum of input-referred noise, in addition to the vacuum noise already present at its input \cite{Caves1982PRD,Clerk2010RMP}. Phase-sensitive detection can avoid that added-noise penalty for one quadrature, whereas photon counting measures a different observable. Thus a claim of performance below the SQL must identify the reference measurement.

Noise acts on both the encoded state and the information reaching the detector. Markovian Lindblad dynamics are appropriate only when memoryless evolution is a justified approximation; colored environments may require a different description \cite{Lindblad1976CMP,deVega2017RMP}. Dephasing and loss can remove ideal entanglement scaling, although useful finite-resource gains can remain \cite{Huelga1997PRL,Escher2011NatPhys,Demkowicz2012NatCommun}. The sensor coherence time also differs from the halo coherence time: protecting the probe does not make an unknown stochastic drive indefinitely phase coherent. These limitations motivate the first enhancement mechanism considered below, which modifies the sensor--environment dynamics themselves.

\subsection{Bound-State-Enhanced Axion Detection}
\label{sec:bound}
A system--environment bound state can retain a nondecaying component of a probe excitation and thereby preserve information that would be lost under a dissipative Markov approximation. This is a spectral mechanism, more restrictive than the presence of environmental memory alone. Non-Markovian metrology has been investigated in several noise models, while bound-state protection has been developed for quantum optical metrology and quantum simulation \cite{Chin2012PRL,An2019PRL,Liu2024PRA}. Dong et al. recently applied this idea to axion-mediated spin interactions \cite{Dong2026Arxiv}. Their work is an October 2026 preprint; the predictions discussed here have not been demonstrated in an axion experiment.

Short-time environmental correlations offer a second way to influence the precision of a spin probe. In a seven-qubit quantum simulation, Long et al. tested entanglement-assisted frequency estimation in engineered colored noise by choosing the interrogation time within the quadratic-decay (Zeno) regime \cite{Long2022PRL}. Dong et al. calculated how a short-time Gaussian noise model changes weak-field amplification in a proposed spin-based axionlike-dark-matter search \cite{Dong2025PRD}. A short-time reduction in decoherence can be used in repeated interrogations. The long-time mechanism considered here requires a discrete eigenmode whose energy remains sensitive to the signal.

The proposed apparatus contains polarized $^{129}$Xe source and sensor ensembles, prepared and read out through alkali-spin coupling. A pseudoscalar neutron--neutron interaction transmitted by virtual axion exchange produces an effective field at the sensor. We denote its coupling product by $\lambda_{nn}=g_{ps}^n g_{ps}^n$. In the geometry assumed in the preprint, the source precession produces $B(t)=B_{\mathrm{ac}}\cos(2\pi\nu t)$, with $B_{\mathrm{ac}}=\lambda_{nn}/\mathcal Q_{\mathrm{geom}}$. The factor $\mathcal Q_{\mathrm{geom}}$, denoted $Q$ in the original work, contains the source polarization, separation, neutron mass, nuclear magnetic moment, and Yukawa-range dependence \cite{Dong2026Arxiv,Su2024PRL}. It is unrelated to a resonator quality factor. The source frequency is controlled experimentally; the mediator mass enters the force range, not a halo Compton-frequency resonance. Accordingly, this is a laboratory force-sensing proposal with possible relevance to axion technology, rather than a demonstrated direct-dark-matter search.

In the effective resonant rotating-wave model adopted by Dong et al., the sensor Hamiltonian is 
\begin{equation}
H_0=\omega_A\sigma_z/2   
\end{equation}
with $\omega_A=\omega_0-\gamma_nB_{\mathrm{ac}}/2$ and $\omega_0=\gamma_n B_0=2\pi\nu$. The same model assumes a zero-temperature bosonic reservoir and an initial equal superposition of the effective spin states. Its survival amplitude $\alpha(t)$ obeys \cite{Dong2026Arxiv}
\begin{equation}
\dot\alpha(t)+i\omega_A\alpha(t)
+\frac14\int_0^t f(t-\tau)\alpha(\tau)\,d\tau=0.
\label{eq:bound_memory}
\end{equation}
under $\alpha(0)=1$. $f(t-\tau) = \int_{0}^{\infty}J(\omega)\exp[-i\omega(t-\tau)]d\omega$ is the environmental correlation function. For its Ohmic-family spectral density,
$J(\omega)=\eta\omega(\omega/\omega_c)^{s-1}e^{-\omega/\omega_c}$, $\eta$ is the dimensionless coupling strength, $\omega_c$ is the cutoff frequency, and $s>0$ is the spectral exponent. These parameters describe sensor noise, not the axion halo spectrum.
The poles of the amplitude satisfy
\begin{equation}
y(\varpi)\equiv\omega_A+\frac14\int_0^\infty
\frac{J(\omega)}{\varpi-\omega}\,d\omega=\varpi .
\label{eq:bound_pole}
\end{equation}
An isolated pole $\varpi_b<0$ exists when $y(0)<0$, which for this spectral family becomes
$\omega_A<\eta\Gamma(s)\omega_c/4$, with $\Gamma$ the Euler gamma function \cite{Dong2026Arxiv}. After the continuum contribution dephases, the surviving amplitude is
\begin{equation}
\alpha(t)\longrightarrow Z e^{-i\varpi_b t}.
\label{eq:bound_residue}
\end{equation}
where $Z=\left[1+\frac14\int_0^\infty
\frac{J(\omega)}{(\omega-\varpi_b)^2}\,d\omega\right]^{-1}$.
The residue $0<Z\leq1$ sets the retained amplitude. Memory effects without such an isolated, signal-sensitive pole need not preserve long-time information.

For the resulting reduced spin state, the single-spin QFI for $B_{\mathrm{ac}}$ is
\begin{equation}
F_{B_{\mathrm{ac}}}(t)
=|\partial_{B_{\mathrm{ac}}}\alpha|^2+
\frac{[\partial_{B_{\mathrm{ac}}}|\alpha|^2]^2}
{4|\alpha|^2(1-|\alpha|^2)} .
\label{eq:bound_qfi}
\end{equation}
This expression follows the Bloch-state QFI, with pure-state limits taken continuously \cite{Zhong2013PRA,Dong2026Arxiv}. In the Markov approximation used for comparison, $\alpha(t)=e^{-[\kappa+i(\omega_A+\Delta)]t}$ and $F^{\mathrm M}_{B_{\mathrm{ac}}}=\gamma_n^2t^2e^{-2\kappa t}/4$, after absorbing the frequency shift $\Delta$ and neglecting the field dependence of $\kappa$. The information therefore peaks at a finite interrogation time and decays at long times. This does not imply that an optimized Markovian experiment cannot detect a signal: repeated finite-time interrogations can still accumulate information.

When a bound state exists, the derivative $\partial_{B_{\mathrm{ac}}}\varpi_b=-\gamma_nZ/2$ gives the asymptotic result \cite{Dong2026Arxiv}
\begin{equation}
F_{B_{\mathrm{ac}}}(t)\simeq
\frac{\gamma_n^2Z^4t^2}{4}
+\frac{2-Z^2}{1-Z^2}(\partial_{B_{\mathrm{ac}}}Z)^2 .
\label{eq:bound_asymptotic}
\end{equation}
The leading term restores the ideal $t^2$ dependence but retains a $Z^4$ penalty. It is therefore recovery of a scaling law, rather than complete recovery of the noiseless prefactor. For $N_{\mathrm{se}}$ independent sensor spins and $M$ independent repetitions, the corresponding leading QCRB scale is
\begin{equation}
\delta B_{\mathrm{ac}}^{Q}\simeq
\frac{2}{\gamma_n Z^2t\sqrt{MN_{\mathrm{se}}}},\qquad
\delta\lambda_{nn}^{Q}=\mathcal Q_{\mathrm{geom}}\delta B_{\mathrm{ac}}^{Q}.
\label{eq:bound_sensitivity}
\end{equation}
The $N_{\mathrm{se}}^{-1/2}$ dependence is that of independent spins; the improvement arises from protected time evolution, not an assumed many-spin entanglement gain. Equation~\ref{eq:bound_sensitivity} is a local precision scale, not a confidence-level exclusion by itself.

The preprint also evaluates transverse-polarization readout, $P(t)=\mathrm{Re}\,\alpha(t)$. In the bound-state regime, $P(t)\simeq Z\cos(\varpi_b t)$, and the envelope of favorable readout working points approaches the QFI sensitivity in the reported simulations \cite{Dong2026Arxiv}. Attainability is thus conditional on interrogation timing and phase control; a retained coherence amplitude does not ensure optimal sensitivity at every time. The measurement must also distinguish the weak source-correlated field from magnetic leakage and readout noise.

\begin{figure}[!htbp]
\centering
\includegraphics[width=9cm]{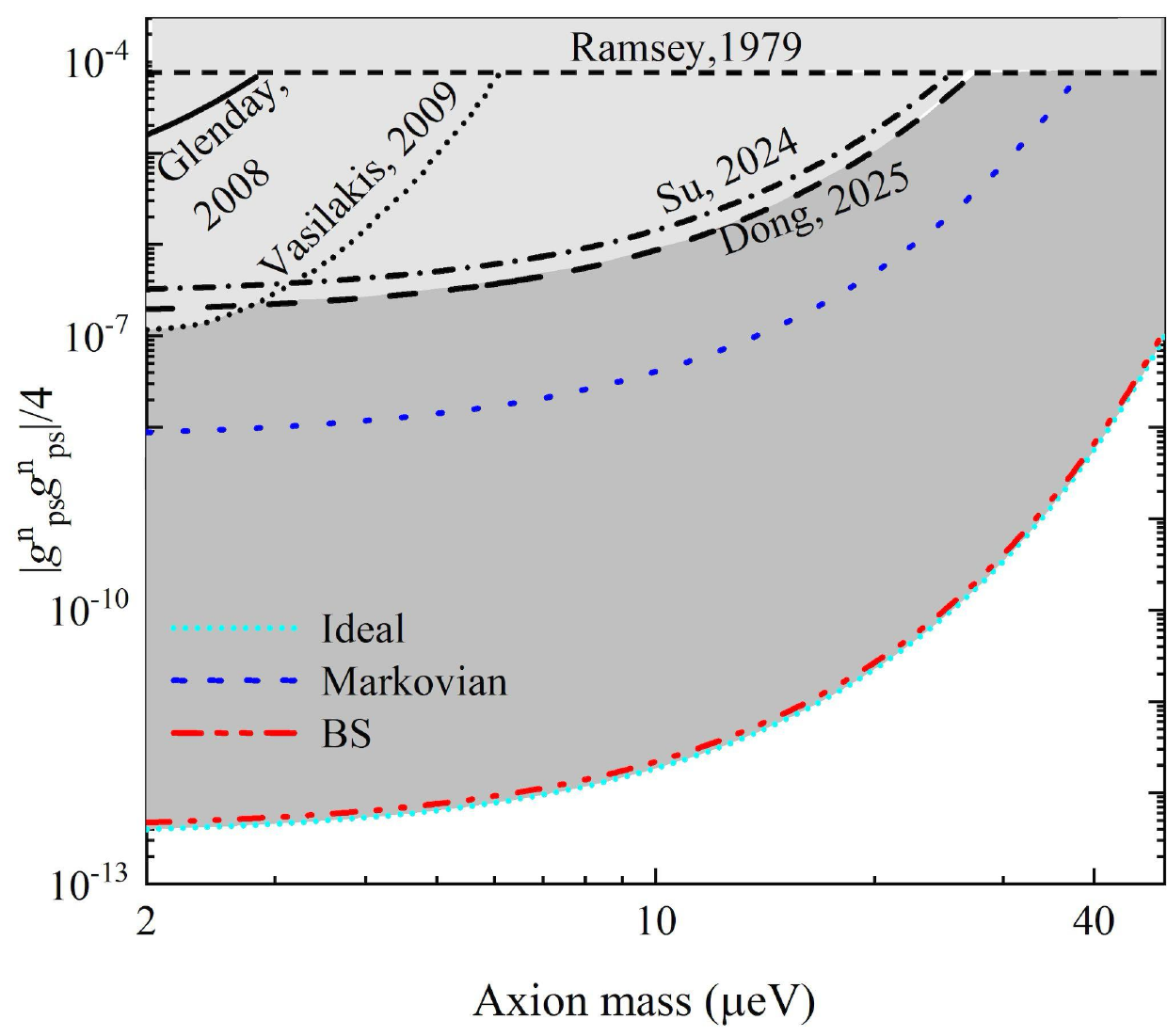}
\caption{Projected sensitivity to the neutron--neutron pseudoscalar coupling product $|g_{ps}^n g_{ps}^n|/4$ in the bound-state model of Dong et al. \cite{Dong2026Arxiv}. The ideal, Markovian, and bound-state curves are theoretical comparisons; the black curves include prior experimental constraints and a theoretical result. The example uses $s=1$, $\eta=0.1$, $\omega_c=50\omega_0$, $t=100/\omega_0$, $M=4.32\times10^4$, and $N_{\mathrm{se}}=2.1\times10^{17}$. It concerns axion-mediated spin interactions, not axion--photon conversion or a measured halo-dark-matter exclusion.}
\label{fig:bound_projection}
\end{figure}

For the numerical example, $s=1$, $\eta=0.1$, $B_{\mathrm{ac}}=140\,\mathrm{aT}$, $N_{\mathrm{se}}=2.1\times10^{17}$, and $M=4.32\times10^4$. The cases $\omega_c=25\omega_0$ and $50\omega_0$ lie on opposite sides of the bound-state threshold, and the coupling comparison in Figure~\ref{fig:bound_projection} uses $\omega_c=50\omega_0$ and $t=100/\omega_0$ \cite{Dong2026Arxiv}. The reported improvement of up to six orders of magnitude near $m_a=3.5\,\mu\mathrm{eV}/c^2$ is a model-based projection relative to previous force measurements, not a newly measured exclusion. It presumes the stated spin number, repetitions, spectral density, and readout model.

Several conditions remain to be established for an implementation. The bath spectrum and occupation must be measured or engineered; a vacuum reservoir cannot simply be identified with the ordinary thermal environment of a low-frequency spin ensemble. The effective rotating-frame Hamiltonian, resonance condition, and reservoir spectrum must describe the same controlled experiment. Independent-spin information addition, residual dephasing, technical noise, and polarization/readout efficiency must also be validated. Finally, keeping $M$ fixed while increasing $t$ increases total exposure; a comparison at fixed total time must include this cost. Bound-state dynamics have been investigated in other platforms, including circuit quantum electrodynamics near a photonic bandgap \cite{Liu2017NP}, but those results do not establish the required reservoir in a xenon force sensor. Extending the proposal to a stochastic halo signal would require a new signal-response analysis with finite field coherence, rather than relabeling its force-sensitivity curve.

\begin{figure}[!htbp]
\centering
\includegraphics[width=16cm]{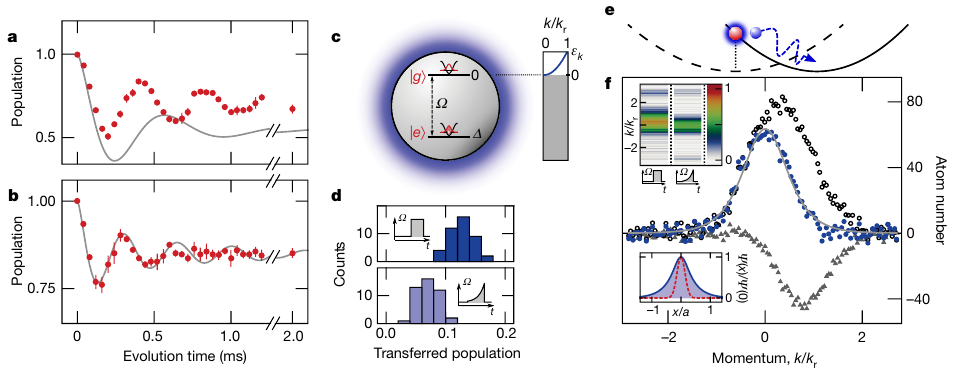}
\caption{Matter-wave bound state in an optical lattice. (a,b) The population of the lattice-confined state oscillates after the coupling is switched on near the continuum edge (a) and at a larger negative detuning (b). (c) The below-threshold dressed state includes an evanescent matter wave. (d) Histograms give the fraction transferred to the other internal state after sudden (upper) and adiabatic (lower) switch-on. (e) An offset axial potential accelerates freely propagating atoms away from the preparation site. (f) Open and filled circles show momentum distributions after sudden and adiabatic preparation, respectively. Triangles show their difference as plotted in the original work. The solid curve is a fit to the adiabatic distribution using the evanescent wavefunction. Reproduced from \href{https://doi.org/10.1038/s41586-018-0348-z}{Fig.~3} of Krinner et al. \cite{Krinner2018Nature}. \textcopyright{} 2018 Springer Nature Limited.}
\label{fig:pub_bound_krinner}
\end{figure}

A tunable matter-wave emitter provides an experimental example of bound-state formation at a continuum edge. Krinner et al. coupled a lattice-confined internal state of $^{87}$Rb to another internal state whose atoms could propagate along a one-dimensional waveguide \cite{Krinner2018Nature}. Figure~\ref{fig:pub_bound_krinner} shows how the preparation protocol and the momentum distribution reveal a localized component.

\subsection{Quantum criticality}
\label{sec:critical}
Critical quantum sensing seeks to increase parameter distinguishability by operating near a quantum phase transition. The rapid change of a many-body or quantum-optical state near criticality can enhance susceptibility and QFI for an appropriate Hamiltonian parameter \cite{Zanardi2008PRA}. Unlike bound-state protection, which preserves information during open-system evolution, criticality changes how strongly the probe state responds to the parameter. A large response is useful only when the associated fluctuations, preparation time, and accessible measurement are included.

In recent years, quantum phase transitions have provided a new approach to improving the sensitivity of weak-signal detection. Near a continuous quantum critical point, the low-energy excitation gap gradually closes, and quantum states and physical observables become highly sensitive to weak external perturbations. Zanardi et al. established a connection between quantum criticality and parameter estimation precision, proposing the use of quantum phase transitions to improve parameter estimation \cite{Zanardi2008PRA}. Subsequently, Chu et al. proposed a quantum metrology scheme based on critical dynamics, enabling high-precision parameter estimation by measuring the dynamical response of a system near a quantum critical point without requiring ground-state preparation \cite{Chu2021PRL}. For a parameter $\theta$ to be estimated, the estimation precision is bounded by the quantum Cramer-Rao bound
\begin{align}
    \delta\theta\geq\frac{1}{\sqrt{M I_\theta}},
\end{align}
where $M$ is the number of independent measurements and $I_\theta$ is the QFI. With appropriate parameter encoding and evolution time, the QFI can be significantly enhanced as the system approaches a quantum critical point, thereby reducing the uncertainty in parameter estimation. Yang and Jacob further proposed a quantum critical detector based on first-order quantum phase transitions, in which weak external signals induce a pronounced macroscopic response in a collective spin system, thereby amplifying the input signals. They also pointed out the potential applications of this approach in microwave single-photon detection and axion detection \cite{Yang2019QE, Yang2019JAP}.

For axions and axion-like particles, their interactions with nuclear spins can be described by an effective weak oscillating pseudo-magnetic field acting on the spin system. Jiang et al. developed a spin-based amplifier using an ensemble of hyperpolarized $^{129}\mathrm{Xe}$ nuclear spins, achieving more than two orders of magnitude amplification of magnetic signals, and applied this technique to the direct search for axion dark matter \cite{Jiang2021NP}.
Subsequently, Su et al. investigated axion-mediated spin interactions using two spatially separated ensembles of $^{129}\mathrm{Xe}$ nuclear spins. The nuclear spins in the source ensemble undergo Larmor precession under an external magnetic field and generate an oscillating pseudo-magnetic field acting on the sensor ensemble through axion-mediated interactions. The experiment exploited spin interactions between Rb and Xe to enhance the detection sensitivity by up to a factor of 145 and established new constraints on the coupling strength in the axion mass range from $3.2$ to $24.3,\mu\mathrm{eV}$ \cite{Su2024PRL}.
Within the effective coherence time of the source spins, one transverse component of the pseudo-magnetic field can be approximated as
\begin{align}
    B_a(t)=B_{\mathrm{ac}}\cos(\omega_dt).,
\end{align}
where $\omega_d$ is the Larmor frequency of the source nuclear spins and $B_{\mathrm{ac}}$ is the amplitude of the pseudo-magnetic field. For the axion-mediated pseudoscalar interaction between neutrons, the pseudo-magnetic field amplitude is related to the axion coupling strength through
\begin{align}
    B_{\mathrm{ac}}=\frac{(g_{\mathrm{ps}}^n)^2}{Q(m_a)}, \label{B_ac}
\end{align}
where $g_{\mathrm{ps}}^n$ is the pseudoscalar coupling constant between axions and neutrons, and $Q(m_a)$ is a coefficient determined by the axion mass $m_a$, the geometry of the experimental setup, and the nuclear spin polarization, among other factors. Therefore, the detection of axion-mediated spin interactions can be converted into the precision measurement of the amplitude of a weak oscillating pseudo-magnetic field.

\begin{figure}[!htbp]
\centering
\includegraphics[width=16cm]{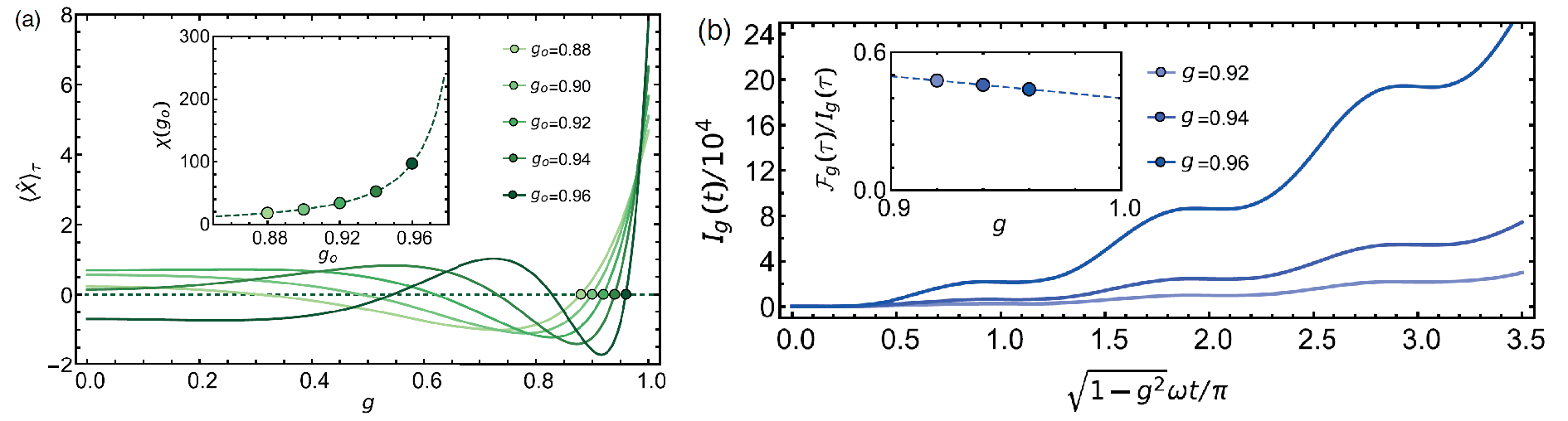}
\caption{Criticality-enhanced quantum sensing based on the quantum Rabi model. (a) Expectation value of the bosonic field quadrature $\langle \hat{X}\rangle$ as a function of the coupling parameter $g$ for different working points $g_0$. The inset shows the enhanced susceptibility $\chi(g_0)$ as the system approaches the quantum critical point. (b) The QFI $I_g(t)$ as a function of evolution time for different coupling strengths. The inset shows that the inverted variance $F_g(\tau)$ obtained from homodyne detection reaches the same order as the QFI, demonstrating enhanced measurement precision near the critical point. Figure adapted with permission of Ref. \cite{Chu2021PRL}.}
\label{fig:quantum_critical}
\end{figure}

In critical quantum systems, both the response rate and the QFI for the parameter to be estimated can be significantly enhanced near the quantum critical point as shown in Fig. \ref{fig:quantum_critical}.
In recent years, the application of the Lipkin–Meshkov–Glick (LMG) model to criticality-enhanced quantum metrology has been extensively investigated \cite{Salvatori2014PRA}.  Luo et al. realized tunable long-range collective spin interactions through cavity-mediated four-photon processes, providing an experimental basis for constructing collective spin sensors exhibiting quantum phase transitions \cite{Luo2025NP}.
To further improve the detection sensitivity of the pseudo-magnetic field, long-range collective spin interactions can be introduced into the sensor spin ensemble to construct a LMG model exhibiting a quantum phase transition. In our group's recent work \cite{Chen2026submitting}, we considered an ensemble of $N$ spin-$1/2$ particles, the effective Hamiltonian can be written as
\begin{align}
    H=-hJ_z-\frac{\lambda_xJ_x^2+\lambda_yJ_y^2}{N}-\gamma_n B_\mathrm{ac}J_y,
\end{align}
where $J_\alpha=\sum_{i=1}^{N}\sigma_i^\alpha/2$ is the collective spin operator, $h$ is the effective Zeeman energy, $\lambda_x$ and $\lambda_y$ are the collective spin coupling strengths, and $\gamma_n$ is the gyromagnetic ratio of the nuclear spins. By tuning the collective spin coupling strengths to bring the system close to a quantum critical point, the enhanced spin response in the critical region can be exploited to improve the detection sensitivity to weak pseudo-magnetic fields. 
By measuring the transverse component $J_x$ of the collective spin, the uncertainty in the pseudo-magnetic field amplitude can be expressed using the error propagation formula as
\begin{align}
    \delta B_{\mathrm{ac}}=\frac{\sqrt{\langle J_x^2(t)\rangle-\langle J_x(t)\rangle^2}}{\sqrt{M}\left|\partial_{B_{\mathrm{ac}}}\langle J_x(t)\rangle\right|}.
\end{align}
Reducing the uncertainty in the pseudo-magnetic field measurement through quantum criticality can further lower the minimum detectable pseudo-magnetic field. Therefore, according to Eq. \eqref{B_ac}, he minimum detectable coupling strength between axions and neutrons satisfies
\begin{align}
    g_{\mathrm{ps}}^n|_{\min}=\sqrt{Q(m_a)B_{\mathrm{ac,min}}}.
\end{align}
If quantum criticality reduces the minimum detectable pseudo-magnetic field by a factor of $K$, the detection threshold for the pseudoscalar coupling constant can be reduced by a factor of $\sqrt{K}$, while that for the product of the coupling constants can be reduced by a factor of $K$. Therefore, combining criticality-enhanced quantum metrology with existing nuclear-spin-based axion detection techniques may further improve the exclusion limits on axion coupling strengths.

\subsection{Squeezing-Enhanced Quantum Sensing}
\label{sec:squeezing}
A small displacement of a microwave mode can be read out as a shift of one field quadrature. For a mode with annihilation operator $b$, Write 
\begin{equation}
 X=(b+b^\dagger)/\sqrt{2},\qquad P=(b-b^\dagger)/(i\sqrt{2})   
\end{equation} 
 Then $[X,P]=i$, and each vacuum variance is $1/2$. An ideal pure state squeezed along $X$ has $(\Delta X)^2=e^{-2r}/2$ and $(\Delta P)^2=e^{2r}/2$, where $r\geq0$. For a displacement of known direction along $X$, the signal shifts the mean of $X$ by the same amount as for vacuum, while its single-shot uncertainty falls by $e^{-r}$. Caves identified this redistribution of fluctuations between conjugate observables as a means of improving interferometric readout \cite{Caves1981PRD}. In a haloscope, the halo phase varies between coherence intervals. A fixed-quadrature measurement samples a projection of the drive, and its coupling sensitivity must be obtained from the distribution of records over those intervals \cite{Brady2022PRXQuantum,Shi2023NPJ}.

 The analogous spin measurement estimates a small rotation $\phi$ of an ensemble initially polarized along $x$. For $N$ spin-$1/2$ particles, let $J_y$ be the transverse spin component read out after a rotation about $z$. At $\phi=0$, the slope of $\langle J_y\rangle$ is $\langle J_x\rangle$. The metrological squeezing parameter is then $\xi_R^2=N(\Delta J_y)^2/|\langle J_x\rangle|^2$, giving $\Delta\phi=\xi_R/\sqrt{N}$ for a single interrogation with ideal spin readout \cite{Wineland1992PRA}. A coherent spin state has $\xi_R^2=1$. Collective spin interactions can shear the fluctuations into an ellipse with a reduced variance along the measured axis \cite{Kitagawa1993PRA}. Retaining polarization is essential because it sets the fringe slope. In an atomic-clock experiment, Hosten et al. measured $20.1\pm0.3$ dB of spectroscopic squeezing and $10.5\pm0.3$ dB of enhancement in the clock measurement, whose performance was limited by microwave phase noise \cite{Hosten2016Nature}. For an axion-induced spin rotation, the gain depends on how much of the measured noise is spin projection noise, on the phase response to the drive, and on spin coherence during interrogation \cite{Aybas2021QST,Kuenstner2026NP}.

Two-mode squeezing prepares correlations in joint quadratures. With suitable phase conventions, $X_1-X_2$ and $P_1+P_2$ commute, and both variances fall below their two-vacuum values in an ideal two-mode squeezed state. Each mode taken separately can have a large variance. Optical measurements by Ou et al. demonstrated Einstein--Podolsky--Rosen (EPR) correlations by using one beam to infer the quadratures of its partner \cite{Ou1992PRL}. This distinction matters for an axion-driven resonator. In the energy-constrained bosonic channel studied by Shi and Zhuang, the random signal appears as excess noise in a probe mode. A retained ancillary mode supplies a reference, and joint processing followed by photon counting can extract information that a single homodyne record loses \cite{Shi2023NPJ}. The comparison assumes a specified signal channel and access to the ancillary output. Entanglement determines which correlations remain available across these subsystems. 

\subsection{Entanglement-Enhanced Quantum Sensing}
\label{sec:states}
When the same field rotates several spins within one coherence region, a many-body superposition can encode their common phase in a collective observable. Consider $N$ spin-$1/2$ probes with generator $J_z=\sum_j\sigma_z^{(j)}/2$ and interrogation time $t$. A product of equatorial spins has $\mathrm{Var}(J_z)=N/4$. A Greenberger--Horne--Zeilinger (GHZ) superposition of all spins up and all spins down has $\mathrm{Var}(J_z)=N^2/4$. From Section~\ref{sec:estimation}, the respective frequency QFIs are $Nt^2$ and $N^2t^2$ for a common unitary rotation \cite{Giovannetti2006PRL,Pezze2018RMP}. The GHZ phase is visible in joint parity statistics, which can be assembled from local outcomes after suitable rotations. These ideal scalings assume a common phase and fixed interrogation time. Under independent dephasing the collective coherence can decay faster as $N$ increases, changing the achievable gain at fixed total exposure \cite{Huelga1997PRL,Demkowicz2012NatCommun}. A steeper phase fringe also has a shorter period. For a single $N$-spin GHZ state, phases separated by $2\pi/N$ give the same parity signal. Cao et al.\ prepared optical-clock GHZ states with up to nine atoms and combined states of different sizes to resolve this ambiguity \cite{Cao2024Nature}. An oscillating halo field presents the additional problem that the phase can change between coherence intervals. A protocol must specify which measurements refer to the same field realization before assigning the gain of a narrow fringe to a coupling estimate.

A distributed sensor can instead be designed to measure a weighted combination of local fields. Entanglement between nodes is useful when its correlations reduce the noise of that particular combination. Xia et al. demonstrated this principle in a three-node radio-frequency sensor network, distributing continuous-variable entanglement among optical probes and combining local readouts to estimate properties including the angle of arrival of a field \cite{Xia2020PRL}. For a microwave dark-matter search, Brady et al. proposed coherently combining cavity outputs from within one field-coherence region and matching the distributed squeezed state to the collected mode \cite{Brady2022PRXQuantum}. The comparison with independent cavities must specify the total collection volume, available squeezing, transmission efficiency, and relative phase stability. The classical intersite correlations used to identify a common field in Section~\ref{sec:platforms} provide a separate experimental benchmark.

Nichol et al.\ linked two $^{88}\mathrm{Sr}^{+}$ ions about $2\,\mathrm{m}$ apart and used entanglement to compare their optical transition frequencies. At fixed interrogation time the uncertainty approached a factor of $\sqrt{2}$ below that of independent ions \cite{Nichol2022Nature}. This comparison senses a differential phase. Identical clocks with identical responses would reject a spatially common axion shift in that channel. A halo application therefore requires a specified difference of responses or a measurement of the common phase.

The best entangled probe also depends on which parameters must be inferred. Chen et al. proposed a hidden-photon protocol with an ideal weak-signal excitation probability proportional to $n_q^2\delta^2$ for $n_q$ qubits and a single-qubit encoding angle $\delta$ \cite{Chen2024PRL}. Santoso and Ho numerically optimized state-preparation and measurement circuits for four- and nine-qubit networks \cite{Santoso2025PRD}. An axion-specific extension by Ho models magnetic conversion followed by a coherent drive on superconducting qubits \cite{Ho2026PRD}. In that proposal the unknown dark-matter phase changes the axis of the qubit rotation. Estimating its two Cartesian components avoids the ambiguity of a phase coordinate at vanishing amplitude, and the simulations use joint estimation to reconstruct the coupling. The resulting precision forecasts depend on the assumed transduction, noise, and circuit performance. Multiparameter comparisons require a common readout and the covariance of the inferred quantities \cite{Ragy2016PRA}.

Non-Gaussian probe states provide a different way to distribute correlations. Freiman et al. proposed sharing an $m$-photon Fock state among $N$ cavities, exposing the array to a spatially coherent dark-photon drive and recombining the outputs \cite{Freiman2026PRD}. Relative to one fixed-size cavity, the collected displacement amplitude grows as $\sqrt{N}$, while stimulated excitation from $m$ to $m+1$ contributes a factor of $m+1$ to the transition probability. Their ideal thermal-background-limited scan rate scales as $N^2(m+1)$ under the specified measurement model. Increasing $N$ at fixed cavity size also increases total collection volume. Quantifying the contribution of entanglement requires a reference array with the same volume, state energy, and loss. For real receivers, the further question is how much of the encoded information reaches the readout port. Section~\ref{sec:receivers} examines that question for squeezed microwave fields.

\subsection{Squeezed-State Receivers and Quantum-Limited Readout}
\label{sec:receivers}
A squeezed microwave receiver uses the quadrature resource of Section~\ref{sec:squeezing} at two distinct points in the measurement chain. A parametric amplifier prepares squeezed vacuum incident on the cavity measurement port. A second amplifier, with its pump phase locked to the first, amplifies the selected output quadrature for detection \cite{Zheng2016Arxiv,Malnou2020PRX,Backes2021Nature}. The resonator adds the axion-conversion field with a phase set by the halo. Averaging over that phase leaves signal power in the measured quadrature. In an overcoupled resonator the reduced incident-quadrature noise maintains sensitivity farther from resonance, permitting a wider frequency step at a given target sensitivity. Internal dissipation introduces noise through a separate port.

Malnou et al. measured $4.5\pm0.1\,\mathrm{dB}$ of squeezing and a scan-rate enhancement of $2.12\pm0.08$ for injected weak classical signals \cite{Malnou2020PRX}. HAYSTAC implemented this architecture in a magnetic haloscope, approximately doubling the search rate and setting limits on the axion--photon coupling \cite{Backes2021Nature}. The subsequent Phase~II analysis reported exclusions in additional mass intervals with the squeezed receiver \cite{Bai2025PRL}. The first measurement characterized the receiver with a known input.           The latter analyses tested halo signals with a calibrated conversion volume, magnetic field, and frequency-dependent exposure.

Transmission loss mixes in environmental noise and degrades squeezing; phase errors mix the antisqueezed quadrature into the measured one. Internal resonator loss also differs from coupling to the readout port, which controls signal extraction and bandwidth \cite{Clerk2010RMP,Chaudhuri2019Arxiv}. Increasing squeezing at its source is therefore insufficient if the useful state does not reach the sensing mode and detector. Receiver optimization must include the measured noise spectrum and the full signal transfer function \cite{Kuo2025PRD}.

A second architecture couples the signal mode to an auxiliary readout mode. The CEASEFIRE proposal combines frequency conversion (state swapping) and two-mode squeezing. Balancing their rates transfers a signal quadrature to the readout while preserving that signal quadrature in the ideal interaction \cite{Wurtz2021PRXquantum}. Jiang et al. inferred a scan-rate ratio of $5.69\pm0.12$ from the measured visibility spectrum and obtained $5.61\pm0.09$ in a search for an injected tone. With a small imbalance and stronger interactions they inferred $8.17\pm0.16$ from the visibility spectrum \cite{Jiang2023PRXQuantum}. These benchmarks quantify the receiver response to synthetic signals. Coupled resonators and squeezed cavity--magnon systems extend this approach in proposals whose forecasts depend on stability, internal damping, and thermal occupation \cite{Chen2022PRR,Ren2026AQT}.

These architectures amplify a field quadrature. When the thermal population and the false-event rate are sufficiently small, individual microwave excitations provide another observable for the same conversion process.

\subsection{Single-Photon Detection}
\label{sec:counting}
Photon counting measures occupation rather than both field quadratures and is therefore not constrained by the phase-insensitive amplifier's added-noise bound. In the low-occupation regime, a weak axion-conversion signal can be sought as excess counts above thermal photons and detector events \cite{Lamoreaux2013PRD}. The relevant limitations are detection efficiency, dark counts, bandwidth, dead time, reset cost, and residual thermal occupation. Counting is not universally superior: a warm or lossy detector may lose its advantage, and a count record generally discards field-phase information that a coherent receiver could use \cite{Kuo2025PRD}.

A dispersively coupled qubit can repeatedly interrogate the photon number of a storage mode. Dixit et al. used this quantum-nondemolition readout in a dark-photon search \cite{Dixit2021PRL}. A counter for traveling photons can instead use a driven dissipative process to convert an incoming photon into a persistent qubit excitation. Lescanne et al. demonstrated this mechanism and measured both its detection efficiency and its dark-count rate \cite{Lescanne2020PRX}. Coupling a counter to an axion haloscope further requires transporting photons from a magnetic conversion volume to a superconducting detector with sufficiently low loss.

The readout cycle must also fit within the lifetime of a stored excitation. Chang et al. used detuning-robust control pulses and repeated parity measurements to read out a short-lived microwave mode. They obtained a dark-photon constraint with no applied conversion field \cite{Chang2026PRD}. For higher photon frequencies, Devlin et al. proposed using cyclotron absorption by a single trapped electron to shift the phase of its axial motion. Their projected 30--60 GHz operation would address an axion converter in a different mass band \cite{Devlin2026PRR}. The latter sensitivity depends on photon delivery, capture probability, and axial phase resolution.

Braggio et al. integrated a transmon-based single-microwave-photon counter with an axion haloscope \cite{Braggio2025PRX}. The search over 30.477--30.479 $\mu\mathrm{eV}/c^2$ reported a 95\% confidence-level upper limit on $|g_{a\gamma}|$ of approximately $7\times10^{-14}\,\mathrm{GeV}^{-1}$ and a factor of about 20 in scan-rate advantage relative to the quantum-limited linear-amplifier reference used in that work. These quantities characterize a particular apparatus and comparison, not a universal counter advantage. 

\begin{figure}[!htbp]
\centering
\includegraphics[width=12cm]{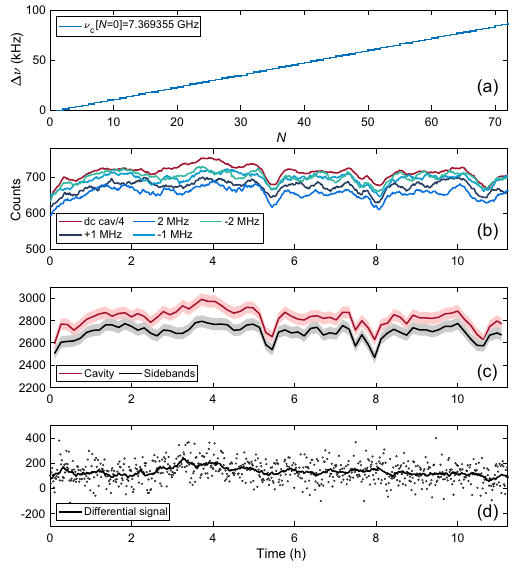}
\caption{Photon counts during a haloscope scan from $7.369355$ to $7.369442\,\mathrm{GHz}$. (a) Cavity tuning through 72 cycles. (b) Counts in four sidebands, offset by $\pm1$ and $\pm2\,\mathrm{MHz}$, compared with one quarter of the on-resonance count $N_c$. Each sideband contributes $7.15\,\mathrm{s}$ of counting time per cycle. (c) The sideband sum $N_b$ and $N_c$ for equal live times of $28.6\,\mathrm{s}$. Shading marks one Poisson standard deviation. (d) The difference $N_c-N_b$. Reproduced from \href{https://doi.org/10.1103/PhysRevX.15.021031}{Fig.~3} of Braggio et al.\ \cite{Braggio2025PRX} under \href{https://creativecommons.org/licenses/by/4.0/}{CC BY 4.0}.}
\label{fig:published_count_background}
\end{figure}

A transmon records individual conversion candidates together with photons from the microwave environment. In the haloscope of Braggio et al., the counter alternated between cavity resonance and four nearby frequencies to estimate the background during each tuning step. Figure~\ref{fig:published_count_background} shows how this procedure acts on a representative count record \cite{Braggio2025PRX}.

The sideband subtraction suppresses slow fluctuations common to both records. A residual excess persisted across cavity tunings, and the authors included a bias term for a small thermal imbalance in extracting the coupling limit \cite{Braggio2025PRX}. An earlier calibration of this counter architecture measured a detection efficiency of $0.43$ and an operational dark-count rate of $85\,\mathrm{s}^{-1}$, dominated by residual photons in the input line \cite{Balembois2024PRApplied}.

A linear-amplifier search provides a complementary reference. The QUAX search around $10.2\,\mathrm{GHz}$ used a traveling-wave parametric amplifier and obtained limits at the $10^{-14}\,\mathrm{GeV}^{-1}$ level \cite{Infirri2025PRL}. The two experiments illustrate distinct readout choices; their quoted coupling limits alone cannot rank the receivers because frequency, volume, conversion field, exposure, and noise differ.

State preparation can also change how many signal-induced transitions a counter records. Agrawal et al. prepared microwave photon-number states to stimulate transitions from a weak dark-photon drive \cite{Agrawal2024PRL}. Zheng et al. used four-component cat states in a dark-photon search and reported an 8.1-fold enhancement of the signal photon rate \cite{Zheng2026PRL}. The coupling reach depends additionally on background events, preparation fidelity, loss, and repetition rate. Transferring these protocols to axions would require a specified magnetic conversion stage and an end-to-end efficiency measurement.

Across these examples, the quantum resource enters a different part of the measurement. Reservoir engineering can preserve an encoded phase, a correlated probe can increase the information acquired during interrogation, and a receiver determines which information reaches the recorded data. Section~\ref{sec:outlook} considers which calibrations and experimental tests are needed to translate these gains into coupling reach over a specified mass range.

\section{Conclusion and Outlook}
\label{sec:outlook}
Axion searches connect a specific particle interaction to a calibrated measurement. Photon conversion, spin-gradient coupling, and oscillating EDMs lead to different signal observables, while axion-induced clock shifts require their own atomic and nuclear response models. Distributed and hybrid platforms expand frequency coverage and background discrimination, but their use of quantum degrees of freedom does not alone establish an improvement over a conventional benchmark.

In a magnetic haloscope, squeezing can broaden the useful search band and photon counting can improve sensitivity at low thermal occupation. Both readouts have been used in axion searches \cite{Backes2021Nature,Braggio2025PRX}. Other developments address different stages of the measurement: spin control and hybrid transduction improve access to weak drives, while nonclassical cavity states and distributed resources offer new ways to encode or combine information. Their gains must be evaluated after loss, stochastic field coherence, calibration uncertainty, and operational overhead, with scalar and dark-photon results retained under their actual interaction models.

Bound-state protection and criticality identify complementary theoretical opportunities. The former can preserve signal-dependent phase in a structured environment, but the axion-related proposal reviewed here concerns a controlled source--sensor force and retains a finite-residue penalty. The latter can enhance parameter distinguishability, but preparation time, finite-frequency response, and fluctuations prevent a direct identification of critical susceptibility with dark-matter sensitivity. For both routes, an experimentally characterized noise model and an accessible readout are prerequisites for a credible coupling forecast.

Some of these mechanisms can be combined. A cavity network can distribute squeezed fields and combine outputs when its relative phases track the same halo realization \cite{Zhang2021QST,Brady2022PRXQuantum}.A protected spin could also be paired with a lower-noise readout, provided the protection preserves the signal generator and the readout still resolves it. The gains share noise sources and observing time. Their combined value must be calculated from the joint response and noise covariance or measured with matched calibration signals.

The published spectra suggest a practical test of a proposed improvement. In a spin or dielectric detector, measure the response to a calibrated drive and the output noise over the same band. For a microwave receiver, propagate both through the analysis that sets a coupling limit or scan rate. A larger conversion peak, greater visibility away from resonance, and a lower counting background then have distinct, measurable consequences \cite{Xu2024CP,Ary2025PRL,Braggio2025PRX}.

Improved sensitivity also raises the standard for identifying a candidate. Ahn et al. investigated an excess near 1.036 GHz using independent checks and the original apparatus. The excess did not persist, and the extended scan set coupling limits over 1.026--1.045 GHz \cite{Ahn2026PRL}. A search should therefore report how calibration, rescans, and the trials factor enter a detection claim alongside the noise and bandwidth of the receiver.

Quantum error correction illustrates the same distinction at the level of probe dynamics. Under Markovian noise and sufficiently fast controls with noiseless ancillas, restoring Heisenberg scaling in interrogation time requires the signal Hamiltonian to have a component outside the Lindblad span generated by the noise \cite{Zhou2018NC}. A logical-memory benchmark tests a different task \cite{Acharya2025Nature}. Likewise, the entanglement-assisted trajectory discrimination studied by Chin et al. calls for a different signal model from halo-field estimation \cite{Chin2025PRL}. A credible forecast follows the axion interaction, the noise it competes with, and the measurement that preserves the distinction all the way to the final likelihood \cite{Sushkov2023PRXQuantum,chou2023arXiv,HAZUMI2026NIMPR}.



\medskip
\textbf{Acknowledgements} \par 
This work is supported by Guangdong Provincial Quantum Science Strategic Initiative under Grant No.~GDZX2505004, the National Natural Science Foundation of China under Grant No.~62461160263, and Quantum Science and Technology-National Science and Technology Major Project (2023ZD0300200).

\medskip

%
\bibliographystyle{MSP}
\bibliography{ref}

\end{document}